\documentclass[elsart,12pt]{article}
\usepackage{graphicx}
\usepackage{epsfig}
\begin{document}
\baselineskip 20pt plus .1pt  minus .1pt
\pagestyle{plain}
\voffset -2.0cm
\hoffset -1.0cm
\setcounter{page}{01}
\rightline {}
\vskip 1.0cm

\begin{center}
{\Large {\bf PROPERTIES OF THE INTERSTELLAR MEDIUM AND THE
PROPAGATION OF COSMIC RAYS IN THE GALAXY}}
\end{center}
\begin{center}
A.D.Erlykin \footnote{Corresponding author: Erlykin A.D.,
e-mail: A.D.Erlykin@durham.ac.uk}$^{,2}$, A.A.Lagutin $^3$,
A.W.Wolfendale $^1$ \end{center}
\begin{flushleft}
(1) Department of Physics, University of Durham, South Road,
Durham     DH1 3LE, UK \\
(2) P. N. Lebedev Physical Institute, Leninsky Prosp., Moscow
    117924, Russia \\
(3) Altai State University, Dimitrova st. 66, Barnaul 656099,
Russia \\ 
\end{flushleft}

\begin{abstract}
The problem of the origin of cosmic rays in the shocks produced
by supernova explosions at energies below the so-called 'knee' 
(~at $\sim 3 \cdot 10^6$ GeV~) in the energy spectrum is
addressed, with special attention to the propagation of the
particles through the inhomogeneous interstellar medium and the
need to explain recent anisotropy results, \cite{Tsune}. It is
shown that the fractal character of the matter density and
magnetic field distribution leads to the likelihood of a
substantial increase of spatial fluctuations in the cosmic ray
energy spectra. While the
spatial distribution of cosmic rays in the vicinity of their sources 
(~{\em eg.} inside the Galactic disk~) does not depend much on the character
of propagation and is largely determined by the distribution of
their sources, the distribution at large distances from the
Galactic disk depends strongly on the character of the
propagation. In particular, the fractal character of the ISM
leads to what is known as 'anomalous diffusion' and such
diffusion helps us to understand the formation of the Cosmic
Ray Halo. Anomalous diffusion allows an explanation of the
recent important result from the Chacaltaya extensive air
shower experiment \cite{Tsune}, viz. a Galactic Plane
Enhancement of cosmic ray intensity in the Outer Galaxy, which
is otherwise absent for the case of the so-called 'normal'
diffusion.  All these effects are for just one reason: anomalous
diffusion emphasizes the role of local phenomena in the
formation of cosmic ray characteristics in our Galaxy and
elsewhere.  
\end{abstract}
{\Large {\bf 1. Introduction}}
Since the early fifties supernova remnants (~SNR~) have been
considered as the most likely sources of Galactic cosmic rays
(~following the pioneering work of I.S.Shklovsky~). 
The main arguments for his proposal were sufficient energy
deposition and the presence of non-thermal radio emission which
indicates that SNR are the sites of electron acceleration. Later
theoretical analysis of diffusive shock acceleration, which was
supposed to be the main mechanism of particle acceleration in
SNR, confirmed that they can produce a power law spectrum of
cosmic rays over a wide range of energies, and thus explain one
of the most prominent features of the cosmic radiation. 

Although there is evidence at GeV gamma ray energies for SNR
acceleration of the initiating particles \cite{Bhat,Osbor},
since the development of high energy (~TeV~) gamma ray
astronomy doubts have begun to arise as to the role of SNR as
the main sources of cosmic rays above some 10's of GeV. These
doubts are inspired by the continuing absence of direct
observations of TeV gamma-rays from known SNR which could be
associated with the accelerated protons and nuclei. Another
serious problem for the SNR model, and, indeed, for any model in
which the source distribution is similar to that of SN, in terms
of dependence of number on Galactocentric distance, is that the
radial gradient of the proton and nuclei intensity in our Galaxy
has been found to be less than the gradient of supernovae. 
Interestingly, the gradient for electrons - as determined from
low energy gamma rays \cite{Issa}, and synchrotron radiation
\cite{Broa1} - seems 'normal'. 

Further problems are that the energy of the knee (~$\sim10^6$
GeV~) is above the maximum energy of accelerated protons in all
models. Also, the required $\geq$10\% efficiency of the energy
transfer from the kinetic energy of the SNR shock to the energy
of accelerated particles may well be too high in that the
efficiency observed hitherto appears to be less than about 1\%
\cite{Plaga}. However, it must be said that the problem of
maximum energy is less severe if, as we maintain, the knee is
associated with 'heavy nuclei', specifically oxygen and iron
\cite{EW2,EW3}. A way forward is to search for other sources of
TeV cosmic ray nuclei (~{\em eg} \cite{Plaga}~) or to re-examine
the manner in which cosmic rays propagate through the Galaxy,
viz, whether or not the propagation has the usually-adopted
Gaussian form (~spatially, about the source~). It is the latter
approach that we adopt here. Specifically, we wish to see to
what extent, the adoption of a more realistic model of cosmic
ray diffusion in the ISM helps in the quest for an explanation
of the following 'cosmic ray anomalies': 
the small cosmic ray gradient (~for protons and nuclei~), 
the extended cosmic ray halo,
the Galactic Plane Enhancement in the Outer Galaxy and the
energetics problem. \
However, indirectly we also test the basic conception that
the main source of cosmic rays are SNR with their diffusive
shock acceleration mechanism producing the energy spectrum with
a power law exponent about 2.15.

{\Large {\bf 2. The properties of the interstellar medium (~ISM~)}} 
{\large {\bf 2.1. Connection of the diffusion characteristics
with the spectrum of magnetic turbulence in the ISM}}

It is well known that the distribution of matter and magnetic
fields in the Galaxy is highly non-homogeneous on different
spatial scales. Gaseous clouds with very different densities,
temperatures and degrees of ionization move through space in a
highly turbulent way. Many evidences both from theory and
observations of the existence of multiscale structures in the
Galaxy have been found during a few last decades (~see, for
instance, 
\cite{Lee,Kapla,Lozin,Ruzma,Vains,Bochk,Falga,Burla,Molch,
Meish,Armst,Elme1,Minte,Cadav,Elme2}.
Filaments, shells, clouds and other entities are widely spread in
the ISM. A rich variety of structures which are created in
interacting phases and have different properties can be related
to the fundamental property of the turbulence, called the intermittency.

In 1941 A.V.Kolmogorov, assuming that all
the characteristics of the turbulent medium depend {\em only} on
the energy flux through the turbulent vortex hierarchy, derived
an expression for the spectrum of turbulence (~the energy per
unit mass and wave number bin, $k$~). For steady-state
hydrodynamic conditions, when the medium is incompressable,
magnetic fields are absent and the energy flux does not depend
on the viscosity, the spectrum is 
$F(k) \propto k^{-\chi}$ with $\chi = \frac{5}{3}$ \cite{Kolmo}.
At this point it is commonly assumed that not only is the
Kolmogorov expression valid for cosmic ray propagation, but,
furthermore, that the diffusion process is 'normal', viz.
Gaussian. Specifically, for one-dimensional propagation we can
write, for the particle density, $\rho$, at distance $x$ from a
source and at time $t$ after a burst of particles is emitted: 
\begin{equation}
\rho(x,t) = \frac{1}{\sqrt{2\pi
\sigma}}exp(-\frac{x^2}{2\sigma^2}) 
\end{equation}
where $\sigma = \sqrt{2Dt}$, $D$ being the diffusion
coefficient. It will be shown
that neither assumption is valid. Concerning the value of
$\chi$, this changes if magnetic fields are present.  

In the case of the gas being conductive, which is the usual state of
the interstellar plasma, the turbulent motion generates magnetic
fields and they, themselves, have a turbulent structure. The
stretching, bending and folding of magnetic field lines by
turbulent motions of the medium partially coupled to the field
make the magnetic field also highly intermittent, especially on
the smaller scales.

Because magnetic irregularities scatter the charged particles, making
the propagation of cosmic rays diffusion-like, the spectrum of
these irregularities is important for the magnitude of the
diffusion coefficient, which governs the transport of the cosmic
rays. In 1965 R.H.Kraichnan showed that for the hydromagnetic
case with the presence of magnetic fields the turbulence
spectrum displays an exact equipartition between the magnetic
and kinetic energies and its shape is modified. The
predicted magnetic and kinetic energy spectra are both
proportional to $k^{-\chi}$ with $\chi = \frac{3}{2}$ instead of
$\frac{5}{3}$ \cite{Kraic}.

As remarked already, many cosmic ray workers have used the
Kolmogorov form - which has become a sort of 'standard' for
cosmic ray diffusion; however, it is now appreciated that,
insofar as most of the ISM is ionized to a significant degree,
the Kraichnan form is more appropriate. In fact, we ourselves
have prefered the latter \cite{EW1} for an empirical reason,
viz. that, when married with a 'standard' SNR production
spectrum, the ensuing ambient cosmic ray spectrum has the
'correct' form.      

The Galactic magnetic field is the superposition of a regular
spiral component $B_0$ and a random component $B_1$, which is
determined by turbulent motions in the ISM. The mean free path
$\lambda_{sc}$ for the scattering of a
particle with gyroradius
$r_g$ by random magnetic irregularities with wave number $k$ is
determined \cite{Longe, Berez} by \begin{equation}
\lambda_{sc} = \frac{r_g B^2_0}{B^2_1(k)k}
\end{equation}
Here $B^2_1(k)$ is the magnetic energy density contained in
irregularities with wave number $k$ and, since it is determined
by the ISM turbulence, $B^2_1(k) \propto k^{-\chi}$. Scattering
of the particles is most efficient in the 'resonant' condition
when their gyroradii, determined mostly by the regular magnetic
field $B_0$, is close to the size of the irregularities, i.e. 
$r_g \approx 1/k$. In this case the transverse diffusion
across the regular magnetic field, which is important for the
radial propagation of cosmic rays in the Galaxy, is
characterized by the diffusion coefficient $D_\perp$, where
$D_\perp$ depends on the energy $E$ as 
\begin{equation}
D_\perp(E) = D_0 E^{\delta}
\end{equation}
$E$ nearly coincides with the rigidity for relativistic protons
and electrons and 
\begin{equation}
\delta = 2 - \chi
\end{equation}
The parallel diffusion {\em along} the magnetic field lines
which is typical for the cosmic ray propagation orthogonal to
the Galactic disk (~there being evidence for field lines
perpendicular to the Galactic Plane in the Halo~), is
usually faster and goes with the parallel diffusion coefficient 
$D_\parallel \geq D_\perp$ \cite{Casse}.

It is important to point out that the simple form for the energy
dependence of the diffusion coefficient (~equation (3)~) can
only apply below a particular energy, $E_0$, usually identified
with that of the knee in the spectrum (~at $\sim$ 3 PeV~). At
higher energies where the spectrum steepens, it is usually
considered that there is insufficient power in the
irregularities to maintain the original value of $\delta$. The
Kolmogorov case corresponds to the assumption of energy
injection (~from SNR, etc.~) at the appropriate distance scale
for $E_0$, viz. a few {\em pc}. At higher energies and bigger
distance scales not only is the form of $D(E)$ very uncertain
but near-straight line propagation is not far off.  

{\large {\bf 2.2. Anomalous diffusion in the fractal ISM}}

Most calculations for the propagation of cosmic rays in the
Galaxy have been
made using 'quasilinear theory' \cite{Jokip}, whose validity is
limited to very low levels of turbulence and a quasi-homogeneous
ISM. The turbulence and diffusion characteristics were connected
only via the energy dependence of the diffusion coefficient
determined by equations (3) and (4). In fact, as we mentioned in
\S2.1, the ISM has the multiphase character and is highly
non-homogeneous. The turbulence level is quite high and the mean
amplitude of the irregular magnetic field in the Galaxy
determined by the turbulence is
generally of the same order of magnitude as that of the regular
field: $\frac{B_1}{B_0} \approx 1.5$ \cite{Broa2} and this ratio
undoubtedly varies from place to place, depending on the
proximity of stars of various types and 'shocks' of a wide range
of strengths. The way forward - to advance on the simple
homogeneous diffusion approximation - is frought with difficulty
but we consider that one, hopefully fruitful, approach is by way
of the so-called 'anomalous diffusion' scenario
\cite{Bouch,Isich,West,Uchai}.  Here there are
two regimes: that of sub-diffusion, in which there
are relatively small spatial displacements and of
superdiffusion, where there are large displacements (~the so
called 'Levy flights'~). 

An approach to anomalous diffusive propagation has been recently
developed by Lagutin et al. \cite{Lagu1,Lagu2}. In this work
the power law character of the turbulence spectrum provides a
self-similarity of the ISM structure, i.e. {\em the ISM is
fractal-like} \cite{Feder}.  This self-similarity is not
global and does not cover the whole Galaxy, but is valid on a
wide range of spatial scales, at least in the pc and sub-pc scales of 
importance here. Indeed the experimental
study of the ISM by radio observations revealed that various
structures of the ISM are not distributed at random, but obey
power-law relations between size, linewidth and mass
\cite{Combe}. These power laws demonstrate the self-similar
nature of the ISM and are observed at the scales ranging from
100 pc down to $2 \cdot 10^{-3}$ pc, i.e. at nearly 5 decades of
the wave number $k$ from $2 \cdot 10^{-20}$ to $10^{-15}$
cm$^{-1}$. It can be that the range is even wider, because
direct measurements within the solar system made by Mariner-4
spacecraft found that above $k \approx 6 \cdot 10^{-12}$
cm$^{-1}$ the wave number spectrum of the magnetic
irregularities is proportional to $k^{-\frac{3}{2}}$ for another
two decades \cite{Longe}. 

On the basis of the assumption that the ISM has a fractal structure Lagutin et
al. have formulated and solved analytically the equation for
anomalous diffusion in a fractal medium for different input
conditions. 
The basic equation for cosmic ray concentration {\em n} without
energy losses and nuclear interactions has the form
\begin{equation}
\frac{dn}{dt} = -D(E,\alpha)(-\Delta)^{\frac{\alpha}{2}}
n(r,t,E) + S(r,t,E)
\end{equation}
where $D(E,\alpha)$ is the anomalous diffusion coefficient,
$\alpha$ is an exponent determined by the fractal structure of
the medium, 
$(-\Delta)^{\frac{\alpha}{2}}$ is the fractional
laplacian (~called 'Riss' operator \cite{Samko}~). We remind
that the normal diffusion is described by the similar equation, 
where the first term in its right part is $D(E) \Delta
n(r,t,E)$, $\Delta$ is an ordinary laplacian and the diffusion
coefficient $D$ depends only on the energy. The details of the
solution of the equation (5) are given in \cite{Lagu1,Lagu2}.

The solution depends on the anomalous diffusion
coefficient $D(E,\alpha)$ and the index $\alpha$
determines the temporal dependence of the cosmic ray propagation
and the shape of the diffusion front. Later we show that its
magnitude can be connected with the spectrum of magnetic
irregularities. Therefore this index is the
fundamental parameter for the propagation of cosmic rays in the
turbulent ISM, which determines and interconnects the energy
dependence and the speed of propagation as well as the shape of 
the diffusion front.

The standard diffusion in a homogeneous medium which leads to a
Gaussian distribution of particle densities corresponds to
$\alpha = 2$. The case of $\alpha < 2$ corresponds to the
so-called superdiffusive regime of anomalous diffusion.

It is appreciated that there is a certain {\em ad hoc} element
in the treatment but, at least, the actual ingredients of the
real ISM seem to be in place.

{\Large {\bf 3. The model}}

Our model for particle acceleration by a SNR shock and
the subsequent propagation through the ISM is actually a
generalization of that described in \cite{EW1}. The supernova 
explosions are assumed to be randomly distributed in time. Their
spatial distribution in the Galactic disk is assumed
cylindrically symmetric, as
\begin{equation}
\rho_{SN} =Af(R,Z)
\end{equation} 
where
\begin{equation}
f(R,Z)=(\frac{R}{R_\odot})^{a}exp(-b(\frac{R}{R_\odot}-1)-\frac{Z}{H_z})
\end{equation}
with $a=1.69\pm0.22, b=3.33\pm0.37, R_\odot=8.5kpc, H_z=0.2kpc$
\cite{Pohl}. The normalization
constant $A$ has been taken to provide the total rate of 
$10^{-2} year^{-1}$ Type II SN explosions in the Galaxy.

The acceleration and the propagation of cosmic rays during the
SN explosion consists of two phases: the expansion and the
diffusion. Details of the original model are given in
\cite{EW1}. 
In the present model the expansion phase of the explosion has
not been changed and at
the end of it the accelerated protons have a nearly power law
energy spectrum with an exponent of about 2.2. 

The diffusion phase of the propagation has been generalized to
study the consequences of the different
turbulence spectra. We have made calculations for 4 values of
$\alpha$: 2, 1.5, 1 and 0.5. The important generalizations were
related to the constraint: 
\begin{equation}
(D(E,\alpha) \tau)^{\frac{1}{\alpha}} = H_G
\end{equation}
and to the expression for the mean radius of the diffusive
front: 
\begin{equation}
R_d = (D(E,\alpha)t)^{\frac{1}{\alpha}} =
H_G(\frac{t}{\tau})^{\frac{1}{\alpha}} 
\end{equation}
Here $\tau$ is the mean lifetime of
particles in the Galaxy which is determined by escape and $H_G$
is the vertical scale parameter of the Galaxy, which we adopt as
1 kpc \cite{EW1}. It is seen from (7) that the normal Gaussian
case arises if $\alpha = 2$ and that the temporal spread of
the particles in the important cases of $\alpha < 2$ is faster
then for 'normal' diffusion. 

On the basis of a correspondence of Richardson's
relative-diffusion law \cite{Richa}: $R_d \propto
t^{\frac{3}{2}}$ to the diffusion in the medium with
Kolmogorov-type turbulence spectrum \cite{Batch,Lesie} we assume
that 
\begin{equation}
\alpha = 2(2-\chi) = 2\delta
\end{equation}
and by this way connect the properties of the turbulent ISM with
the characteristics of the cosmic ray propagation through it.
The shape of the diffusion front is different for different
$\alpha$ and is shown in Figure 1. It is seen that for $\alpha <
2$ there is a tail at large distances due to particles which
propagate with small scattering (~the 'Levy flights' referred to
earlier~).  

As for 'normal' diffusion the normalization of the particle
density profile at any instant of time and in the absence of
escape losses has been made to preserve the total energy
contained in the cosmic rays equal to $10^{50}$ erg. The escape
time has been determined by the constraint (8) and has a power
law energy dependence with the same but negative exponent as
$D(E,\alpha)$, i.e. 
\begin{equation}
\tau = \tau_0 E^{-(2-\chi)}
\end{equation}
with $\tau_0 = \frac{H_G^\alpha}{D_0}$.

{\Large {\bf 4. Energy spectra of cosmic rays}}

The results for the energy spectra of cosmic rays calculated by
us for different values of $\alpha$ are shown in Figure 2. 50
identical samples of 50000 SNR were simulated for 4 values of
$\alpha: 2, 1.5, 1$ and $0.5$.
Better agreement between the observed and simulated spectra
can be achieved by correcting the observed spectrum for
the solar modulation effect at energies below 10 GeV. A higher
absolute intensity in the simulations could be easily obtained
using the higher SNR rate in the nearby Gould Belt and a higher
efficiency for the energy transfer from kinetic energy of the
explosion into accelerated cosmic rays \cite{EW1}.
It is seen that
the mean spectra have different slopes, a result that arises
because we connected the energy dependence of the escape time
$\tau$ and the diffusion  coefficient $D$ with the turbulence
structure of the magnetic field in
the ISM. The more uniform is
the size distribution of magnetic irregularities the weaker is
that dependence and the closer is the spectrum to that emerging
after the acceleration at the end of the SNR expansion
phase.

We remark that the closest to the observed spectrum of
cosmic rays at earth is that with $\alpha = 1.0$ which
corresponds to the turbulence spectrum derived by Kraichnan
\cite{Kraic}, a fact already referred to (~see \S2.1~). Also,
among 50 different samples
there is a good fraction which have
a shape close to that observed below the knee, with its
smoothly steepening background at TeV energies followed by a
concave structure at higher energies. Such a structure of the
spectrum has been explained by us in the Single Source Model
\cite{EW2,EW3}, where the 'knee' has structure due to a
significant contribution from a local SNR. This situation is
still true in the
case of anomalous diffusion.  

{\Large{\bf 5. Spatial distribution of cosmic rays}}

{\large {\bf 5.1. General remarks}} 

For the cosmic ray intensity at large
distances above and below the Galactic Plane the stochastic
character of the spatial distribution of SNR plays an
insignificant role, because all SNR are concentrated in the
disk itself. Thus, in our calculations of the intensity we have
simply made a numerical integration over the smooth SNR
distribution. The propagation of cosmic rays was calculated
assuming that the turbulent properties of the ISM do not depend
on the vertical (~$Z$~) or radial (~$R$~) distance and can
be described by the same power law as within the Galaxy. In
general this can be justified if, for instance, the
regular and the mean irregular magnetic fields decrease with
distance in such a way that the scattering mean free path (1)
does not change.   

{\large {\bf 5.2. The radial distribution}}

{\bf 5.2.1. General remarks}

In Figure 3a we show the distribution of cosmic rays out to
large radial distances in the Galactic disk for two energies: 1 GeV and
1 TeV. For comparison we show also the adopted distributions of
SNR density normalized to 1 at its maximum.
It is seen that the radial distribution of cosmic rays follows
the radial distribution of SNR at almost all distances and for
all values of $\alpha$, a fact due to the high sensitivity of
cosmic ray intensity to nearby sources. The only exception is
the region near the center of the Galaxy (~$R \leq 2$ kpc~),
where there is a lack of sources and cosmic rays fill in the
gap. The significance of the result will be considered
later. 

{\bf 5.2.2. The proton radial gradient}

The gradient can be defined as $S = \frac{d(lnI)}{dR}$, where I is
the intensity of cosmic rays (~or the density of sources~). In
Figure 3a we see that the radial gradient of cosmic rays is
equal to the gradient of the sources, whatever the value of
$\alpha$. At the site of the solar system, i.e. at R = 8.5 kpc
it is $-0.17 \pm 0.05$ kpc$^{-1}$. Experimental values for
the Outer Galaxy derived by us from the gamma-ray emissivity
profile are (-0.10$\pm$0.02) kpc$^{-1}$ in Quadrant 2 and
(0.0$\pm$0.03) kpc$^{-1}$ in Quadrant 3 \cite{Erly1}, {\em ie}
the observed radial gradient is smaller than that predicted, a
well-known fact \cite{Paul}. However as remarked in
\cite{Paul} the overall emissivity gradient does not
necessarily hold locally. In particular, measurements of the
emissivity in the Perseus arm at 3 kpc from the Sun in the
Outer Galaxy and at the distance of Cepheus and Polaris
flare at 0.25 kpc give a higher gradient: $-0.15 \pm 0.03$
kpc$^{-1}$, which does not contradict the gradient of the SNR
distribution \cite{Digel}. It is at greater galactocentric
distances (~$R \geq 15$ kpc~) that the observed gradient is
inconsistent with that predicted.

The maximum of the
gamma-ray emissivity is at about the same distance, $R \approx
4$ kpc, from the Galactic center as the maximum of the
SNR distribution which favors an SNR-origin of the bulk of the
cosmic rays, or at least sources which have the same spatial
distribution.

Since the amplitude of the anisotropy is $A =
\lambda_{sc}S$ and we estimate from our model that for 1 TeV
$\lambda_{sc} \approx 7.5\cdot10^{-3}$ kpc, then we expect
that the anisotropy amplitude at this energy will be about
$(1.3 \pm 0.4) \cdot 10^{-3}$, the experimental value being
close to it: $(0.8 \pm 0.2) \cdot 10^{-3}$ \cite{Watso}

To this extent, we can conclude that the anomalous
(~superdiffusive~) diffusion mechanism does not help in
elucidating the problem of the low overall radial gradient of
protons and nuclei (~at least in the far Outer Galaxy~) and
their association with SNR without additional assumptions ( we
prefer re-entrant particles from the Halo - see
\cite{ELW}

{\bf 5.2.2. The electron radial gradient}

In contrast to the case for protons and nuclei there is no problem
with th origin of cosmic ray electrons in SNR at least at
energies below about some 10's of GeV. The radial gradient at
the site of the solar system (~derived from low energy
gamma-rays~) is $-0.25 \pm 0.02$ kpc$^{-1}$, which overlaps
with the SNR gradient within the error bars \cite{Beuer} and
with that predicted for 'cosmic rays' in general. The direct
radio, X-ray and gamma-ray observations of SNR confirm
the presence of accelerated electrons in the SNR shells. The
estimated efficiency of the acceleration is about 1\% which is
much less stringent than the $\sim$10\% required for protons.
The reason for the difference between electrons and protons is
not clear but it is perhaps due to a different radial
dependence of the injection efficiency for the two particles.

{\large {\bf 5.3. The vertical distribution}}

Another situation is for the vertical direction. The
Galactic disk is rather thin and the SNR distribution in this
direction is narrow. The shape of the diffusion front is much
more important in this case and Figure 3b shows that the longer
tails of anomalous diffusion fronts create a much wider
distribution, forming {\em a cosmic ray halo}, which spreads up
to some tens of kpc outside the Galactic disk.

Interestingly, there is some evidence that there is
significant ionized gas at large values of $Z$, specifically
the mean gas density $\langle \rho \rangle \sim
10^{-3}$cm$^{-3}$ at $Z \sim 50$ kpc \cite{Moore} and it has
been suggested that there is a corresponding significant
cosmic ray intensity \cite{WW1}. Its value is not inconsistent
with the value (~$\simeq2 \cdot 10^{-5}$ of the Galactic Plane
intensity~) found for $\alpha = 1$ at $Z \sim 50$ kpc. In fact
such estimates are very approximate insofar as the diffusion
coefficient $D_\parallel$ is presumably greater in the outer
halo than in the disk.

Further relevance of the
$Z$-distribution is considered later.

{\Large {\bf 6.Galactic Plane Enhancement}}

There is another effect
connected with cosmic ray propagation -
the possibility of
observing the Galactic Plane Enhancement (~GPE~) in the
Outer Galaxy. This quantity, $f_e$, was introduced by Wdowczyk
and Wolfendale \cite{WW2} to evaluate the deviation from cosmic
ray isotropy in terms of its Galactic latitude distribution. We
define this quantity in a somewhat different way from that in
\cite{WW2}, by
\begin{equation}
\frac{F(b)}{F_0}=\frac{DgradI(b)}{cF_0}=\frac{2H_G^\alpha(I(b)-I_0)}{c\tau
\Delta^{\alpha-1}I_0}=a+f_ecos^2b 
\end{equation} 
Here $a$ is
the fitting coefficient which should be equal to 1 for
complete isotropy. $F(b)$ is the flux of cosmic rays from
Galactic latitude $b$ and $F_0$ is the flux expected for an
isotropic distribution. We have calculated these fluxes from
the intensities $I_0$ at the solar system and $I(b)$ - at a
distance of $\Delta = 0.1$ kpc from the Sun in the direction of
latitude $b$ and longitude $l = 0^{\circ}$. In Figure 4 the
ratio of these intensities is shown for an energy of 10$^5$
GeV. It is seen that for ordinary diffusion (~$\alpha = 2$~)
the GPE is observed only in the Inner Galaxy (~$b = 0^\circ$~)
and there is no sign of an enhancement in the Outer Galaxy
(~$b = 180^\circ$~).

Decreasing $\alpha$, however, results in an increase of the GPE in terms
of intensities both in the Inner and in the Outer Galaxy. The
origin of this effect is in the much flatter radial
distribution of sources compared with the vertical one. Due
to the high sensitivity of the cosmic rays to the location of
nearby sources, intensities at $b = 0^\circ$ and $180^\circ$
are determined only by the radial gradient of SNR $S_r \approx
-0.19$ kpc$^{-1}$ for all values of $\alpha$ (~see Fig.3a~).
The vertical gradient of SNR is much larger $S_v \approx -5$
kpc$^{-1}$, however, and the intensity at $b = 90^\circ$ is
sensitive to the gradient of cosmic ray intensity $S_v^\alpha$
nearby the disk which is different for different $\alpha$. For
instance, $S_v^2 = -0.59$ kpc$^{-1}$, $S_v^{1.5} = -0.81$
kpc$^{-1}$, $S_v^1 = -1.15$ kpc$^{-1}$ and $S_v^{0.5} = -0.91$
kpc$^{-1}$. Because the gradient is larger for smaller
$\alpha$, the drop of intensity at 0.1 kpc off the disk is
higher and this overcomes the drop of intensity at the same
0.1 kpc towards the Outer Galaxy at $b = 180^\circ$. When one
approaches the direction of $b = 180^\circ$ from $90^\circ$
the intensity starts to rise again and the GPE becomes
visible.

The experimental data on the anisotropy of EAS of
the energy above 10$^5$GeV observed at Chacaltaya shows clear
evidence for the GPE in the Outer Galaxy \cite{Tsune}.
Experimental points fitted by expression (12) give $f_e =
(1.18\pm0.48)\cdot10^{-3}$. The same fitting of the curves
shown in Figure 4 (~for the Outer Galaxy the latitude $b$ in
the expression (12) is replaced by $b-\pi$ in radians~) give
$f_e = (2.45\pm 0.01)\cdot10^{-5}$ for $\alpha = 0.5$,
$(1.02\pm0.01)\cdot10^{-3}$ for $\alpha = 1$
and $(1.28\pm0.03)\cdot10^{-2}$ for $\alpha = 1.5$. As seen in
Figure 4 there is no GPE for $\alpha = 2$ and the formally
determined $f_e$ has a negative value.

When comparing the
experimental and theoreticasl values of GPE one has to keep in
mind the fact that the experimental value has been derived from
the fluxes averaged over wide intervals of Galactic longitude.
However, it should not have a considerable effect: in the
experiment \cite{Tsune} the averaging has been made over a
$90^\circ$ longitude band centered on $l = 180^\circ$.
Our calculations show that the value $I(l,b)-I_0 \propto
cos(l-180^\circ)$ so that the averaging over the interval of
$l = 135^\circ - 225^\circ$ can diminish our $f_e$ by no more
than 10\%. 

The comparison shows that the closest approach to
the experimental value of $f_e$ is again achieved for $\alpha
= 1.0$. However, stochastic variations of the cosmic ray
intensity due to the proximity of nearby sources can spoil this
agreement.

The most important conclusion from this
consideration and Figure 4 is that the propagation in the
non-homogeneous medium gives an explanation of the
observed positive value of Galactic Plane Enhancement in the
Outer Galaxy, whereas the 'normal' diffusion mechanism fails
here.

A general point that can be made here concerns the {\em
shape} of $I(b)$. Our model gives a good fit to the
experimental data and the empirical function (~equation (12)~).
Detailed calculations have not yet been made but it seems
likely that such agreement can be used to give an upper limit
to the $Z$-distribution of the sources. Such arguments will be
useful as a diagnostic for possible source types. At present we
can rule out source distributions with $\langle Z \rangle >$
several kpc and thus 'objects' which have been projected out
into the Halo and large scale acceleration in the Halo itself.

{\Large{\bf 7. Local deviations of the cosmic ray intensity}}

In our previous work \cite{EW1} we emphasized the
point that there is a
high sensitivity of the observed cosmic
ray intensity and its spectral shape to the local spatial and
age distribution of SNR and here we have shown that this
sensitivity rises with decreasing $\alpha$. In Figure 5 we show
examples of the Galactic latitude distributions for random
distributions of SNR simulated within a radius of 0.5 kpc from
the solar system added to the regular distribution from the
more distant SNR taken within 10 kpc radius. There are 12
independent samples of 1000 random supernovae in each sample
and they were simulated using 4 types of ISM with $\alpha = 2,
1.5, 1$ and $0.5$. 

It is seen again that the fluctuations
increase for lower values of $\alpha$ and there are samples
which exhibit a completely different latitude dependence with a
'negative enhancement' or a displaced phase of an enhancement.
Among the latter there are some samples which demonstrate
maxima at  $l > 180^\circ$. There is a south-north asymmetry
observed in the experiment \cite{Tsune} and a small
displacement of the peak intensity (~to $b \sim 200^\circ$~)
and our simulations indicate that these features could have a
local origin.

{\Large{\bf 8. Conclusions}}

We have shown that the structure of the
ISM can determine some of the
important characteristics of the
cosmic rays in our Galaxy: the slope of the energy spectrum and
its fluctuations, the formation of the halo at large distances
from the Galactic disk, the presence of a Galactic Plane
Enhancement and a south-north asymmetry in the Outer
Galaxy. All our simulations have been made within the
framework of the conception that SNR are the sources of cosmic
rays in our Galaxy and diffusive shock acceleration is
the mechanism of particle acceleration at energies below the
knee. We emphasize the importance of the local structure of the
ISM for the formation of most cosmic ray characteristics. A
comparison of the simulated characteristics with experiment
indicates that the fractal structure of ISM with the parameter
$\alpha = 1$ (~Kraichnan spectrum of magnetic irregularities~)
gives local cosmic ray characteristics which are closest to the
experiment.  

Returning to the specific questions referred to
in \S1, the answers are as follows: \\
The radial gradient:
the sole 'new model' does not help, the explanation needs
additional assumptions. \\ 
Energetics: there is no
improvement.\\ 
GPE in the Outer Galaxy: here there {\em is} an
improvement. The observations are consistent with the anomalous
diffusion model but not with 'normal diffusion'.

As a bonus
we have the Halo and the greater ease of explaining
the spectral shape. 

{\Large {\bf Acknowledgements}}

The authors are grateful to the UK's Particle Physics and
Astronomy Research Council, to The Royal Society and to The
Russian Foundation for Basic Research for financial support
and an unknown referee for the useful remarks.

\newpage

{\Large{\bf Captions to figures}}

Figure 1. Shape of the particle density
distribution for different values of the $\alpha$-parameter. The
curves are normalized at $x = 0$, $x$ is the ratio of distance
from the source to the mean distance. The long 'tail' for
anomalous diffusion will be noted. 

Figure 2. Energy spectra of cosmic rays for different values of the
$\alpha$-parameter. 50 samples of 50000 SNR are simulated in
each case. The region between the two thick full lines, denoted
as 'obs', is the observed cosmic ray spectrum \cite{Apana}.

Figure 3. The radial (a) and vertical (b) lateral distribution of the
cosmic ray intensity for energies 1 GeV (~thick lines~) and 1
TeV (~thin lines~) for different values of the
$\alpha$-parameter. Full line: $\alpha =2$, dashed line:
$\alpha = 1.5$, dash-dotted line: $\alpha = 1$ and
dash-3dotted line: $\alpha = 0.5$. Dotted line - SNR
distribution. 

Figure 4. The Galactic latitude distribution of the cosmic ray intensity
for the energy 10$^5$ GeV and different values of the
$\alpha$-parameter. Full line: $\alpha =2$, dashed line:
$\alpha = 1.5$, dash-dotted line: $\alpha = 1$ and dash-3dotted
line: $\alpha$ = 0.5. The longitude $l$ has been taken as
$0^\circ$ for the Inner Galaxy and $180^\circ$ for the Outer
Galaxy. The GPE is an excess at $b = 0^\circ$ and/or
$180^\circ$. 

Figure 5. The galactic latitude distribution of the relative cosmic ray
intensity for energy 1 TeV and different values of
the $\alpha$-parameter: 2 (a), 1.5 (b), 1 (c) and 0.5 (d).
Intensities include contributions from SNR randomly distributed
in space and time within 0.5 kpc from the
Sun.

\end{document}